\documentclass{iaus}
\usepackage{graphicx}

\title[New evidence for halo gas accretion onto disk galaxies] 
{Halo gas accretion onto disk galaxies}

\author[Filippo Fraternali]   
{Filippo Fraternali$^1$}

\affiliation{$^1$Astronomy Department, University of Bologna,
via Ranzani 1, I-40127, Bologna, Italy \\
email: {filippo.fraternali@unibo.it}
}

\pubyear{2008}
\volume{254}  
\pagerange{}
\jname{The Galaxy Disk in Cosmological Context}
\editors{}

\newcommand {\hi} {{\rm H}\,{\small\rm I}}

\newcommand {\kms} {\,{\rm km\,s}^{-1}}

\newcommand {\de}{^{\circ}}
\newcommand {\mo}{\,{M}_\odot}

\newcommand {\moyr}{\,{M_\odot\,\rm yr}^{-1}}

\newcommand{\gsim}{\lower.7ex\hbox{$\;\stackrel{\textstyle>}{\sim}\;$}}
\newcommand{\lsim}{\lower.7ex\hbox{$\;\stackrel{\textstyle<}{\sim}\;$}}

\begin{document}

\maketitle

\begin{abstract}
Studies of the halo gas in the Milky Way and in nearby spiral galaxies show the 
presence of gas complexes that cannot be reconciled with an internal (galactic 
fountain) origin and are direct evidence of gas accretion. 
Estimating gas accretion rates from these features consistently gives values, 
which are one order of magnitude lower than what is needed to feed the star formation. 
I show that this problem can be overcome if most of the accretion is in fact “hidden” 
as it mixes with the galactic fountain material coming from the disk. 
This model not only provides an explanation for the missing gas accretion but 
also reproduces the peculiar kinematics of the halo gas in particular the 
vertical rotation gradient. 
In this view this gradient becomes indirect evidence for gas accretion.

\end{abstract}

\firstsection 

\section{Introduction}

A large amount of fresh gas accretion onto the Milky Way has been advocated 
for decades since the work of \cite{twarog} 
showed the relative constancy of the Star Formation Rate (SFR) in the Galactic disk
over the last 10 Gyrs.
Roughly constant SFR requires constant accretion and this translates
today into a need for gas accretion at a rate of the order $\sim 1 \moyr$.

Since the 1960s, the presence of a large number of 
neutral hydrogen (\hi) clouds filling the sky around us and 
having, on average, negative velocities with respect to the Milky Way disk 
has been recognized (e.g.\ \cite[Hulsbosch 1968]{huls68}).
Such clouds, named High Velocity Clouds (HVCs), given their largely anomalous 
radial velocities (\cite[Wakker \& van Woerden 1997]{wakker97}), 
were immediately regarded as possible evidence for
gas accretion from intergalactic space onto our Galaxy (\cite[Oort 1970]{oort70}).
Alternative interpretations have been proposed, for example that the clouds are produced
by the cooling of material expelled from the disk via a
``galactic fountain'' (\cite[Bregman 1980]{bregman80}) or that they are a
much more distant Local Group population (e.g.\ \cite[Blitz 1999]{blitz99}).

From the time of Oort's suggestion until very recently there were
 two major unknowns about the HVCs: their distances and their metallicities.
It is thanks to a tenacious observational campaign carried out especially over the last 
few years that we now know both distances and 
metallicities for all the major HVCs 
(\cite[Wakker et al.\ 2007, 2008]{wakker07, wakker08}).
The results leave no doubts: the HVCs are located in the Milky Way halo and 
have metallicities of about an order of magnitude lower than the average disk 
ISM metallicity.
These properties make it most likely that this is accreting material falling
onto the Milky Way for the first time.

The HVCs do not comprise the entire halo population, rather
they are an extreme population at large heights and having
particularly anomalous velocities.
In the lower halo large amounts of cold gas are observed (e.g.\ 
\cite[Lockman 1984]{Lockman}),
some of which go under the name of Intermediate Velocity Clouds (IVCs) and are
regarded as a likely galactic fountain component (e.g.\ \cite[Wakker 2001]{Wakker01}).
More recently, \cite{kalberla08} showed that up to 10\% of the Milky Way \hi\ 
gas is in fact extra-planar and highly turbulent.

\section{Halo gas in nearby spiral galaxies}

In the last decade, the study of the halo (extra-planar) 
gas has been extended to nearby galaxies.
This is a demanding task, given the low surface brightness of the halo emission,
and the studies have been restricted to a relatively small number of objects.
In Table \ref{tExtra}, I summarize the results for the best studied galaxies 
so far.
This table includes galaxies seen at different inclination angles. 
For edge-on galaxies the halo gas can be separated spatially from the disk gas, 
whilst for galaxies seen
at intermediate inclinations it can be separated thanks to its peculiar 
kinematics.
The main kinematic feature of extra-planar gas is its decreasing rotation 
velocity with increasing height from the plane, it is said to be ``lagging'' 
behind the disk gas.
Such a velocity gradient has been estimated for a few galaxies (column 9, 
Table \ref{tExtra}) and it is an
important constraint for models of extra-planar gas formation (see Section 4).
The presence of this gradient is also the main reason why extra-planar gas can
be detected in non edge-on galaxies (\cite[Fraternali et al.\ 2002]{frat02}).

\begin{table}[ht]
\caption{Physical properties of extra-planar gas in spiral galaxies}
\begin{center}
\begin{tabular}{lccccccccc}
\hline\noalign{\smallskip}
Galaxy&Type&incl&v$_{\rm flat}$&M$_{\rm HI_{\rm halo}}$&M$_{\rm HI_{\rm tot}}$&SFR&Accr.\ rate&Gradient$^a$&Ref.\\
          &  &{\scriptsize($\de$)}&{\scriptsize(km/s)}&{\scriptsize($10^8 \mo$)}&{\scriptsize($10^9 \mo$)}&{\scriptsize($M_{\odot}$/yr)}&{\scriptsize ($M_{\odot}$/yr)}&{\scriptsize (km/s/kpc)}&\\
\noalign{\smallskip}
Milky Way & Sb  & -       & 220      & $\sim4$    & 4      & $1-3$& $\approx0.2^b$&$-22$& (1,2,3)\\
M\,31     & Sb  & 77      & 226      & $>0.3$     & 3      & 0.35     &  -  & -        & (4,5)\\
NGC\,253  & Sc  & $\sim$75& $\sim$185& 0.8        & 2.5    & $>10$    &  -  & -        & (6)\\
M\,33     & Scd & 55      & 110      & $>0.1$     & 1      & 0.5      &0.05$^c$& -     & (7,8)\\
NGC\,2403 & Scd & 63      & 130      &  3         & 3.2    &   1.3    & 0.1 &$\sim-12$ & (9) \\
NGC\,2613 & Sb  & $\sim$80& $\sim$300& 4.4$^d$    & 8.7    &   5.1    &  -  & -        & (10) \\
NGC\,3044 & Sc  & 84      & 150      & 4          & 3      & $2.6^e$  &  -  & -        & (11) \\
NGC\,4559 & Scd & 67      & 120      & 5.9        & 6.7    & $0.6^e$  &  -  &$\sim-10$ & (12)\\
NGC\,5746 & Sb  & 86      & 310      & $\sim1$    & 9.4    & 1.2      &0.2$^f$&  -     & (13,14) \\
NGC\,5775 & Sb  & 86      & 200      &  -         & 9.1    & $7.7^e$  &  -  & $-8^g$   & (15) \\
NGC\,6946 & Scd & 38      & 175      & $\gsim$2.9 & 6.7    &   2.2    &  -  & -        & (17)\\
NGC\,891  & Sb  & 90      & 230      & 12         & 4.1    &   3.8    & 0.2 & $-15$    & (18) \\
UGC\,7321 & Sd  & 88      & 110      & $\gsim 0.1$& 1.1    & $\sim0.01^h$& -&$\gsim-25$& (19) \\
\noalign{\smallskip}\hline
\end{tabular}
   \label{tExtra}
\end{center}
{\footnotesize
$^a$ Gradient in rotation velocity with height (from the flat part of the rotation curve);
$^b$ from complex C and other clouds with known distances in (2)
without correction for the ionised fraction;
$^c$ from the \hi\ mass in (8) without their correction for the ionised fraction;
$^d$ from sum of the various extra-planar clouds;
$^e$ calculated from the FIR luminosity using the formula in \cite{kewley02};
$^f$ from the counter-rotating cloud (13) using an infall time-scale of $1\times10^8$ yr;
$^g$ calculated using optical lines (16);
$^h$ SFR of only massive stars $> 5 \mo$.
References:
(1) \cite{kalberla08};
(2) \cite{wakker07, wakker08}; 
(3) \cite{levine08};
(4) \cite{thi04};
(5) \cite{walterbos94}; 
(6) \cite{boomsma05};
(7) \cite{reakes78};
(8) \cite{grossi08};
(9) \cite{frat02};
(10) \cite{chavez01};
(11) \cite{lee97};
(12) \cite{barb05}; 
(13) \cite{rand08}
(14) \cite{pedersen06};
(15) \cite{irwin94};
(16) \cite{heald06a};
(17) \cite{boomsma08}; 
(18) \cite{oost07};
(19) \cite{mat03}.
}
\end{table}

\begin{figure}[ht]
\begin{center}
\includegraphics[width=.41\textwidth]{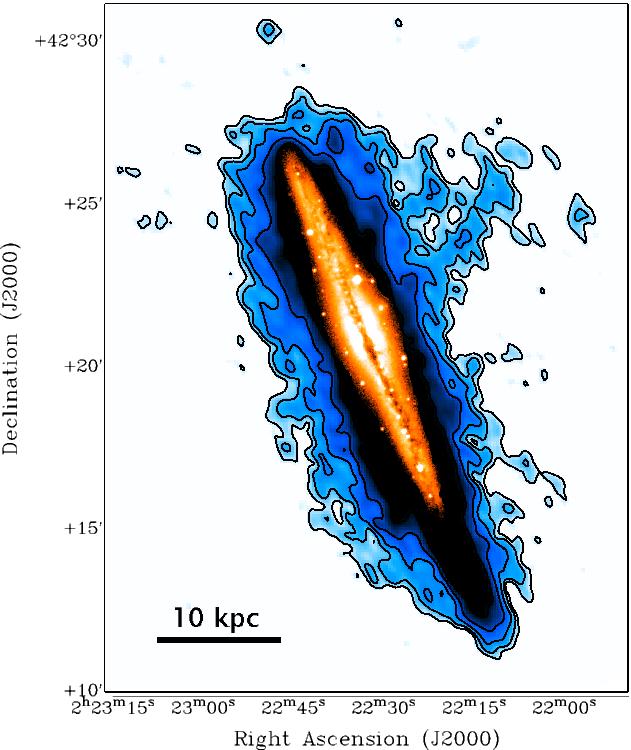} 
\includegraphics[width=.58\textwidth]{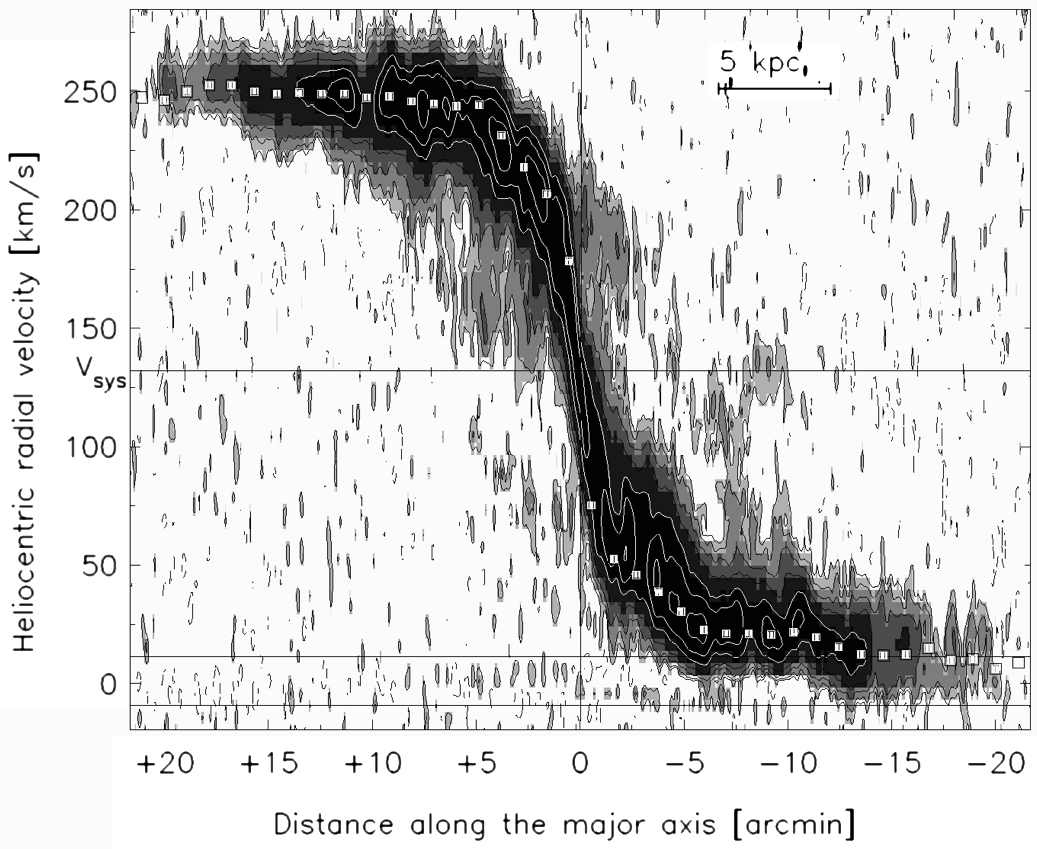} 
 \caption{Two methods of detecting extra-planar gas.
Left: total \hi\ map (blue $+$ contours) for the edge-on galaxy NGC\,891 overlaid on the 
optical image (orange) (data from \cite[Oosterloo et al.\ 2007]{oost07}). 
The extra-planar gas is clearly separated on the sky.
It surrounds the whole galactic disk with a filament that extends up to 20 kpc.
Right: position-velocity diagram along the
major axis of the intermediate inclination galaxy NGC\,2403 (from 
\cite[Fraternali et al.\ 2001]{frat01}). The extra-planar gas is kinematically
separated from the disk gas. It is seen as a faint component 
rotating more slowly than the disk (white dots $=$ disk rotation curve).
}
   \label{891and2403}
\end{center}
\end{figure}

Fig.\ \ref{891and2403} shows the two types of extra-planar gas detections.
On the left, the total \hi\ map of the edge-on galaxy NGC\,891 (blue $+$ contours) is overlaid onto a
DSS optical image (orange).
The \hi\ data (obtained with the Westerbork Synthesis Radio Telescope) show 
a massive and extended \hi\ halo with a 
mass of $1.2 \times 10^9 \mo$ or 30\% of the total \hi\ mass
(\cite[Oosterloo, Fraternali \& Sancisi 2007]{oost07}).
On the right panel of Fig.\ \ref{891and2403}, 
the position-velocity (p-v) plot along the major axis of the galaxy
NGC\,2403 (inclination $60\de$) obtained with the VLA (\cite[Fraternali et al.\ 2001]{frat01}).
A broad component of gas at rotation velocities lower than the disk gas 
(also called the ``beard'') is clearly visible at low emission levels (light grey).
This beard component is the halo gas in NGC\,2403.

Extra-planar gas is possibly ubiquitous as
several nearby galaxies other than those reported in Table \ref{tExtra} show
indications of its presence. It is also observed in the ionised phase.
Optical studies of nearby edge-on galaxies show that roughly half of them have extended layers 
of diffuse ionised gas (\cite[e.g.\ Rossa \& Dettmar 2003]{rossa}) and with 
similar kinematics as the \hi\ layers (\cite[Heald et al.\ 2006b]{heald06b}).

\section{Direct evidence for gas accretion}

Direct evidence of gas accretion is difficult to obtain.
The strategy is to look for gas components (usually at very anomalous velocities)
which are incompatible with an internal origin.
The large majority of the 
extra-planar gas studied so far has actually a very regular 
kinematics that follows closely the kinematics of the disk (see for instance the 
p-v diagram for NGC\,2403, right panel of Fig.\ \ref{891and2403}). This points to
a tight connection between disk and halo components.

\begin{figure}[ht]
\begin{center}
\includegraphics[width=.35\textwidth, angle=0]{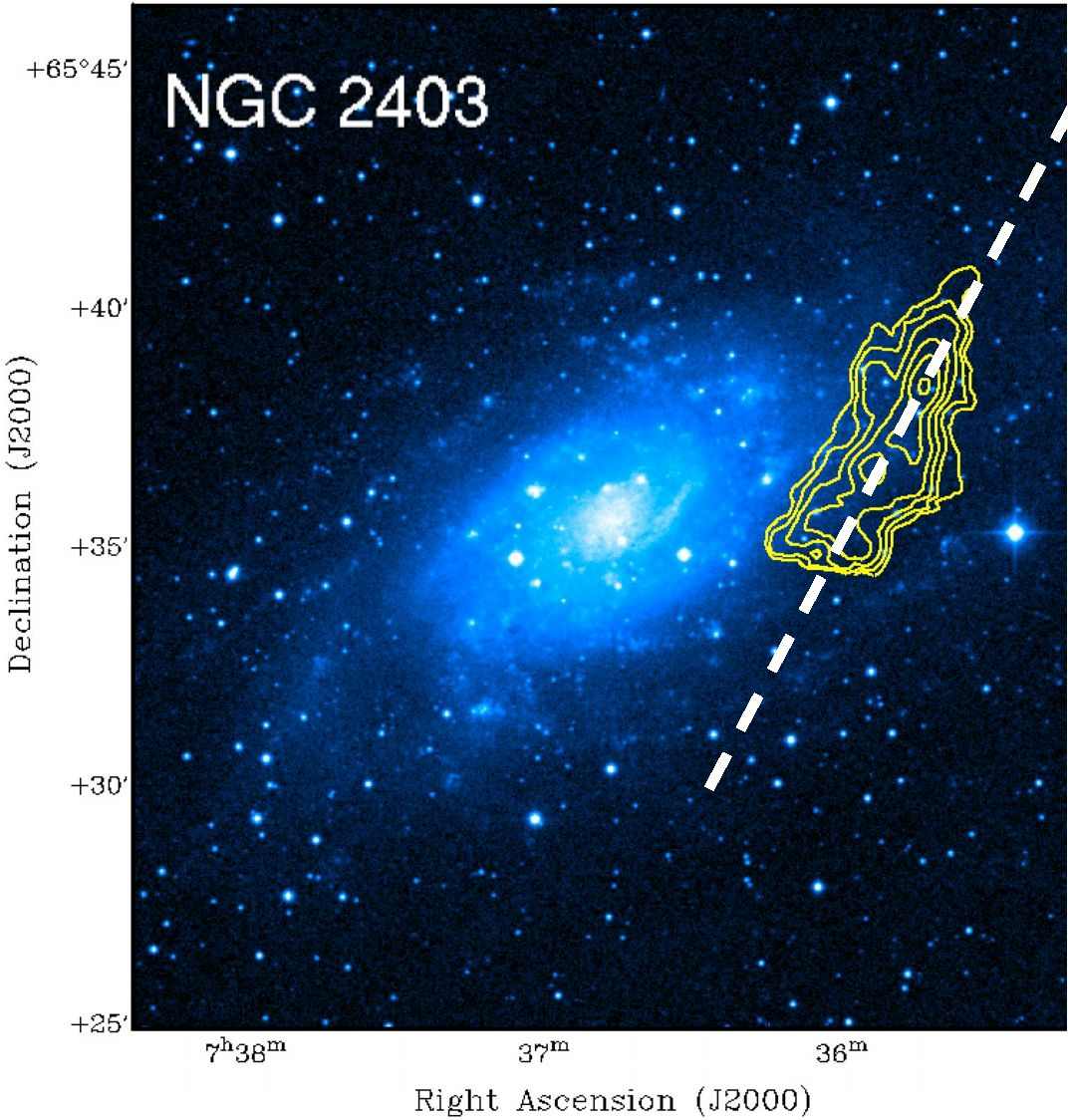} 
\includegraphics[width=.64\textwidth, angle=0]{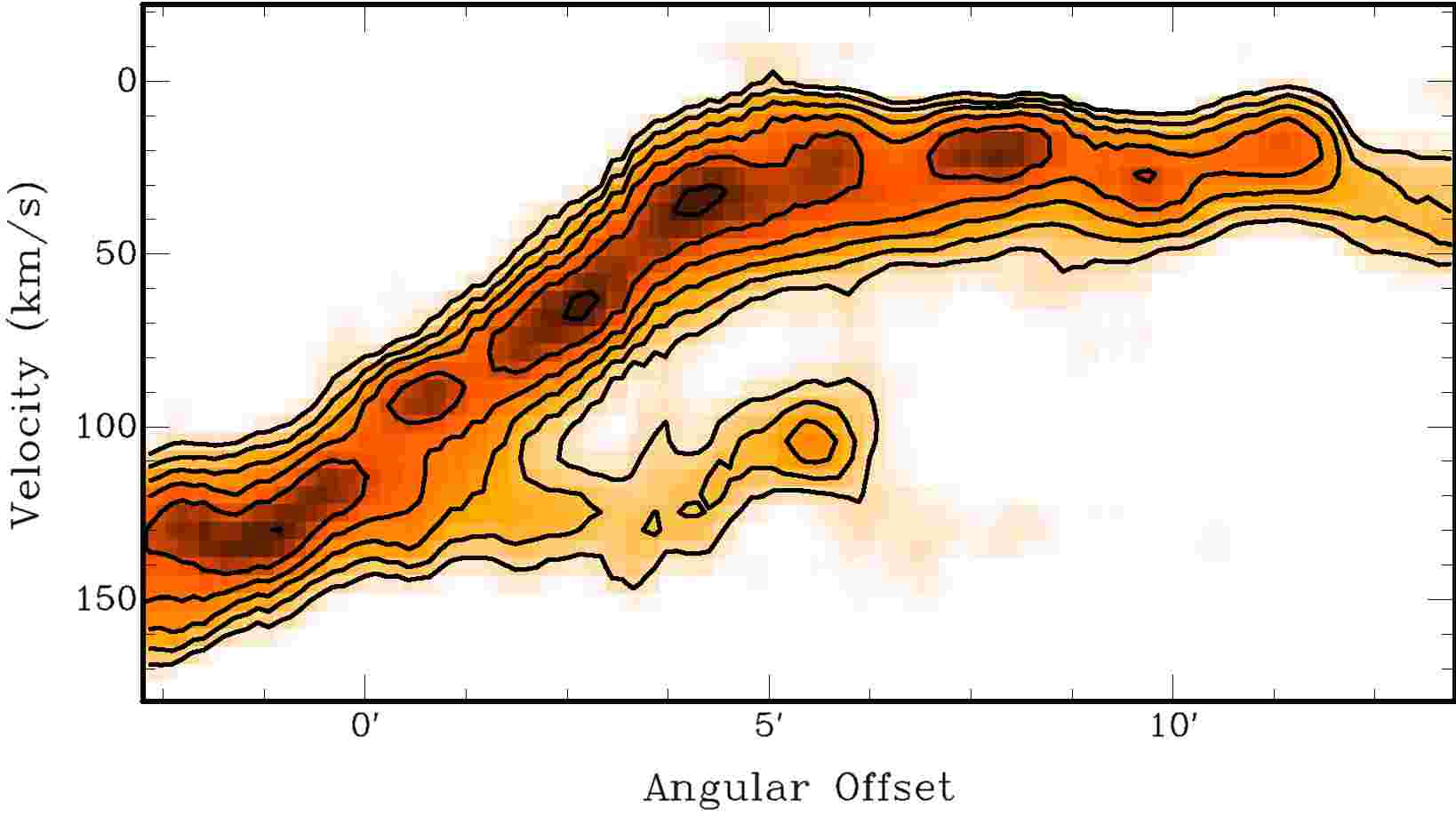} 
 \caption{The \hi\ filament in NGC\,2403. Left: optical image overlaid with the \hi\
map (in contours) of the $1 \times 10^7 \mo$ filament. 
Right: position-velocity plot along the dashed line in the left panel.
The filament is 8\,kpc long and separated from the normal disk kinematics (higher contours) by about $90 \kms$.
Note the small dispersion in velocity of the filament and its proximity to the systemic
velocity of the galaxy ($v_{\rm sys}=133 \kms$).}
   \label{fil2403}
\end{center}
\end{figure}

The first features that deserve attention in the search for gas accretion are 
the \hi\ filaments.
Among the galaxies in Table \ref{tExtra} at least half show large filamentary 
\hi\ structures in their halos,
the most notable cases being NGC\,891 and NGC\,2403.
NGC\,891 has a long massive filament ($M_{\rm HI} \sim 1.6 \times 10^7 \mo$) 
extending up to about 20 kpc from the plane of the disk 
(Fig.\ \ref{891and2403}).
NGC\,2403 also has a filament with a similar \hi\ mass located in projection 
outside the bright optical disk and clearly separated in velocity from the 
disk kinematics (Fig.\ \ref{fil2403}).
They are both very similar to Complex C in our Galaxy.
We can calculate the energy needed to form these filaments
assuming that they come from the disk through a galactic fountain.
In the case of the NGC\,891 filament it turns out that this energy should be
of the order $\sim 1 \times 10^{55}\,$erg.
This would correspond to the explosion of about $10^5$ supernovae in a 
specific region of space and over a time-scale shorter than a dynamical time,
which is clearly a very unlikely event.

\begin{figure}[ht]
\begin{center}
\includegraphics[width=\textwidth]{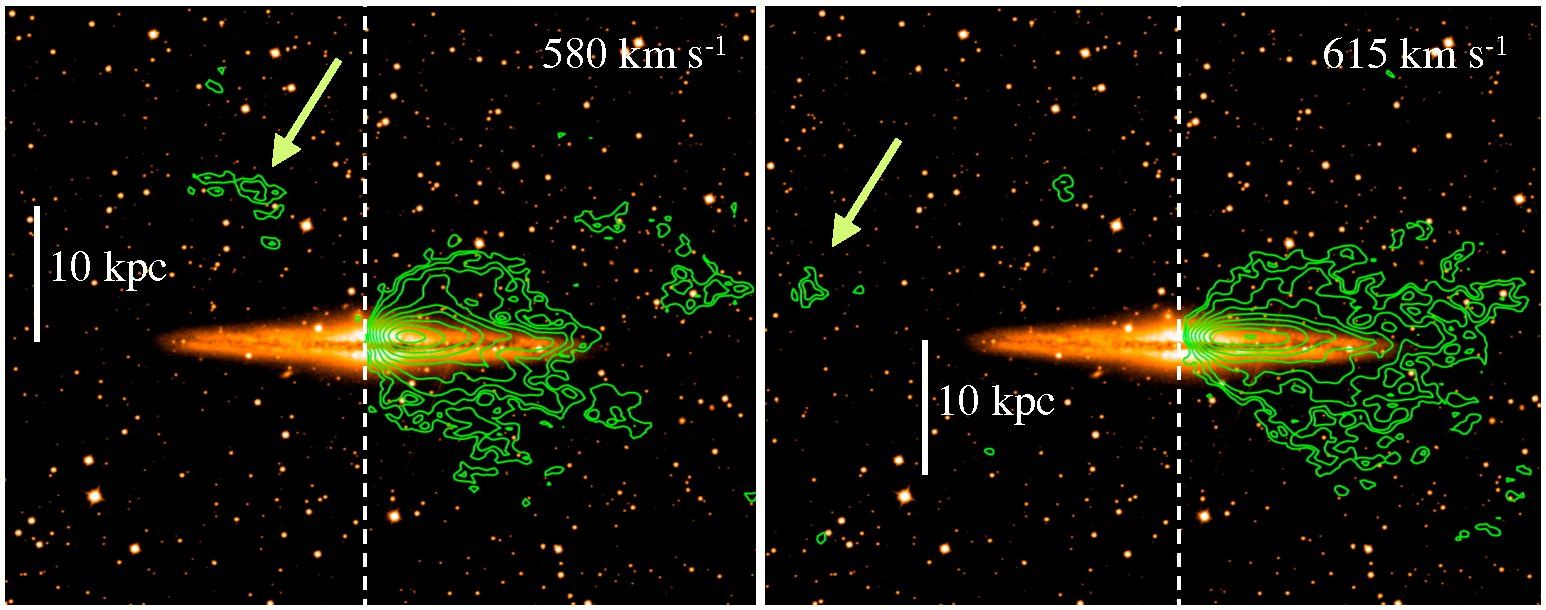} 
 \caption{Counter-rotating clouds in NGC\,891. These two channel maps at radial velocities 
580 and 615 $\kms$ ($v_{\rm sys}=528 \kms$) show gas on the receding side 
(to the right of the vertical dashed line) of the galaxy.
Two clouds are visible in isolation on the left side of the galaxy (arrows), 
they are counter-rotating (from \cite[Oosterloo et al.\ 2007]{oost07}.}
   \label{avcs891}
\end{center}
\end{figure}

A second type of feature that has been found in these new deep surveys are
clouds at very anomalous velocities that end up in the region of 
counter-rotation.
The data of NGC\,891 show two such clouds with masses of order
$\sim 10^6 \mo$ and counter-rotating velocities of $\sim50$ and 
$\sim90 \kms$ (Fig.\ \ref{avcs891}).
These clouds cannot be produced in any kind of galactic fountain and they are
most likely direct evidence of gas accretion.
NGC\,2403 also has gas components at very anomalous velocities called ``forbidden gas''
(Fig.\ \ref{891and2403}; \cite[Fraternali et al.\ 2002]{frat02}). 
Counter-rotating or forbidden clouds are also observed in NGC\,4559,
in NGC\,5746 and NGC\,6946.

If we believe that the above structures have an extragalactic origin we can estimate 
the rate of gas accretion by assuming typical infalling times of ${\rm few} 
\times 10^7-10^8\,$yr.
The resulting rates are shown in Table \ref{tExtra} (column 8), they are typically of 
the order $0.1 \moyr$ and generally 1 order of magnitude lower than the SFRs.
These directly observed accretion rates include only \hi, and they
should be corrected for helium and possibly ionised gas fractions. 
However, it appears difficult to reconcile them with the rates of star 
formation (column 7, Table \ref{tExtra}).
This result is very common for nearby galaxies 
(\cite[Sancisi et al.\ 2008]{sancisi08}).

\section{Indirect evidence for gas accretion}

\begin{figure}[ht]
\begin{center}
\includegraphics[width=.49\textwidth, angle=0]{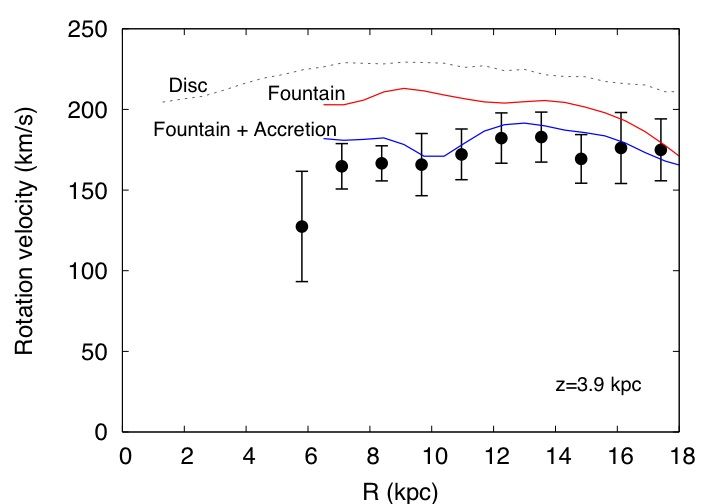} 
\includegraphics[width=.49\textwidth, angle=0]{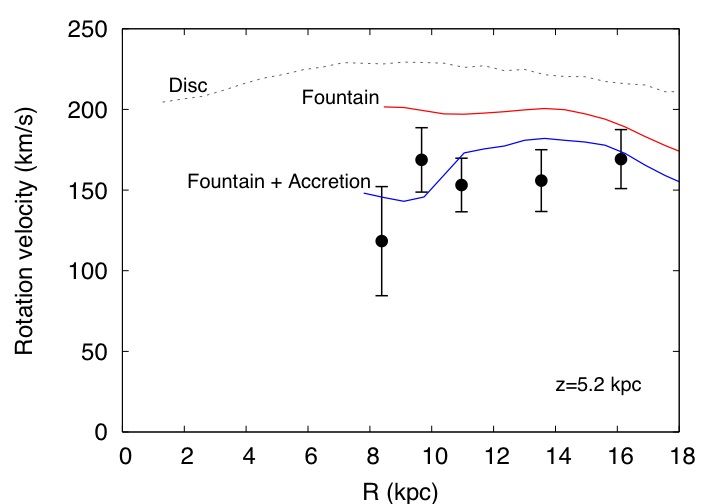} 
 \caption{
Indirect evidence for gas accretion.
Rotation velocities (black points) of the \hi\ extra-planar gas in NGC\,891 at 3.9\,kpc (left)
and 5.2\,kpc (right) from the plane compared to the disk rotation curve (dotted line) and
predictions from two models:
a pure galactic fountain model
(red line above) and a model where the fountain clouds sweep up and accrete ambient gas during their
passage through the halo (blue line below).
The accretion rate required to produce this fit is $\sim 3 \moyr$, very similar
to the star formation rate of NGC\,891
(from \cite[Fraternali \& Binney 2008]{fb08}).
}
   \label{indirect}
\end{center}
\end{figure}

The result that the rate of gas accretion onto galaxies which is 
directly observed is much lower 
than expected implies that most of the accretion should be somewhat ``hidden''.
I describe here possible indirect evidence of this missing gas accretion, 
provided by the rotation velocity gradient of the extra-planar gas.
The steepness of this gradient
is not reproduced by galactic fountain models (e.g.\ \cite[Fraternali \& Binney
 2006; Heald et al.\ 2006b]{fb06, heald06b}) as 
they tend to predict shallower values (a factor half or less).
Fig.\ \ref{indirect} highlights this problem for NGC\,891.
The points are rotation velocities derived at heights $z=3.9\,{\rm kpc}$ and 
$z=5.2\,{\rm kpc}$ from the plane (\cite[Fraternali et al.\ 2005]{frat05}).
Clearly the fountain clouds in the model rotate too fast (have a larger angular momentum)
than the extra-planar gas in the data.

\begin{figure}[ht]
\begin{center}
\includegraphics[width=.65\textwidth, angle=0]{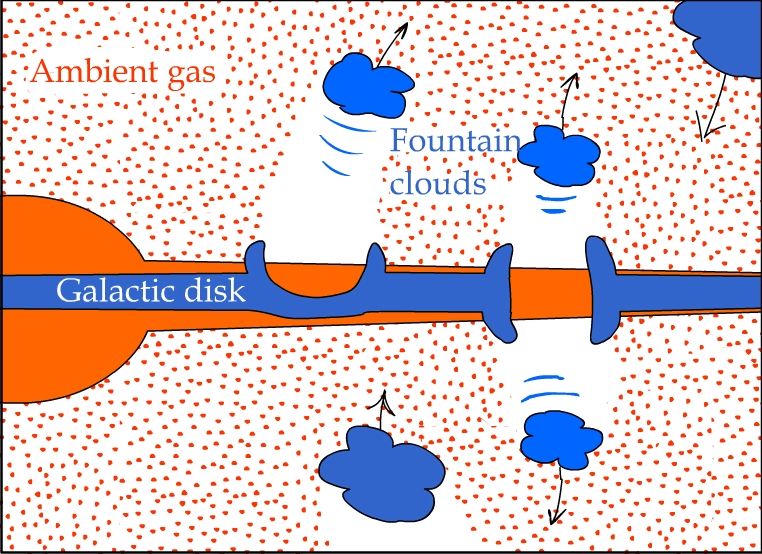} 
 \caption{Schematic of the fountain gas sweeping up ambient gas. The model presented in
\cite{fb08} suggests that superbubbles blow gas into the halo which then sweeps up part of the 
ambient gas during its passage before falling back to the disk.
This mechanism allows fountain clouds to loose part of their angular momentum as
required by the data (see Fig.\ \ref{indirect}) and produces a net gas accretion at
a rate similar to the SFR.}
   \label{cartoon}
\end{center}
\end{figure}

How can the fountain clouds loose part of their angular momentum?
\cite{fb08} consider the possibility that fountain clouds sweep up 
ambient gas as they travel through the halo.
In this scheme ambient gas condenses onto the fountain clouds, these latter grow
along their path through the halo and eventually fall down into the disk 
(see Fig.\ \ref{cartoon}).
If the ambient gas has relatively low angular momentum about the z-axis
then this condensation produces a reduction in the angular momentum of the fountain gas.
The only free parameter of the model is the accretion rate, which is
tuned to reproduce the rotation curves of the extra-planar gas.
Remarkably, the required gas accretion rate turns out to be very similar
to the SFR.
For NGC\,891 we found a best-fit accretion rate of about $3 \moyr$ 
(see the blue curves in Fig.\ \ref{indirect}) and for NGC\,2403: $0.8 \moyr$.
In NGC\,2403, this model is also able to reproduce the observed 
radial inflow of the halo gas (see \cite[Fraternali \& Binney 2008]{fb08}).

One implication of the above fountain$+$accretion model is that
it predicts that most of the extra-planar gas is produced by the galactic
fountain and only a small fraction (about 10\%) is extragalactic.
This is in agreement with the metallicity of the IVCs and the clear links 
between anomalous velocity clouds and star forming regions (e.g.\ 
\cite[Boomsma et al.\ 2008]{boomsma08}).
A second implication is that it does not require that the accreting gas is in 
any particular
phase but only that its angular momentum about the z-axis is 
less than about half the angular momentum of the disk material.
Finally, this model predicts
an accretion rate of the order of the SFR
and in general, proportional to the supernova rate.
Interestingly, this appears to be a general requirement for galaxies throughout
the Hubble time as it reconciles the observed
cosmic star formation history with the gas mass in galaxies at low and high 
redshifts (\cite[Hopkins, McClure-Griffiths \& Gaensler 2008]{hopkins}).


\end{document}